# Fast predicting the complex nonlinear dynamics of mode-locked fiber laser by a recurrent neural network with prior information feeding

*Guoqing Pu, Runmin Liu, Hang Yang, Yongxin Xu, Weisheng Hu, Minglie Hu, and Lilin Yi\**

G. Pu, H. Yang, Y. Xu, W. Hu, L. Yi
State Key Lab of Advanced Communication Systems and Networks, School of Electronic Information and Electrical Engineering, Shanghai Jiao Tong University, Shanghai, 200240, China
E-mail: lilinyi@sjtu.edu.cn

R. Liu, M. Hu
Ultrafast Laser Laboratory, Key Laboratory of Opto-electronic Information Science and Technology of Ministry of Education, College of Precision Instruments and Opto-electronics Engineering, Tianjin University, 300072 Tianjin, China

**Abstract**
As an imperative method of investigating the internal mechanism of femtosecond lasers, traditional femtosecond laser modeling relies on the split-step Fourier method (SSFM) to iteratively resolve the nonlinear Schrödinger equation suffering from the large computation complexity. To realize inverse design and optimization of femtosecond lasers, numerous simulations of mode-locked fiber lasers with different cavity settings are required further highlighting the time-consuming problem induced by the large computation complexity. Here, a recurrent neural network is proposed to realize fast and accurate femtosecond mode-locked fiber laser modeling for the first time. The generalization over different cavity settings is achieved via our proposed prior information feeding method. With the acceleration of GPU, the mean time of the artificial intelligence (AI) model inferring 500 roundtrips is less than 0.1 s. Even on an identical CPU-based hardware platform, the AI model is still 6 times faster than the SSFM method. The proposed AI-enabled method is promising to become a standard approach to femtosecond laser modeling.

## 1. Introduction

Femtosecond mode-locked fiber lasers now occupy the heart position in high-precision metrology including distance ranging,[1] frequency measurement,[2] and various imaging applications.[3][4] The characteristics of femtosecond pulses (i.e., pulse durations and pulse shapes, etc.) generated by mode-locked fiber lasers can vary with different cavity settings and the versatile pulse output from the dedicatedly-designed mode-locked fiber laser is promising to meet the requirements of different applications. Femtosecond mode-locked fiber laser modeling is an imperative approach for theoretically investigating the evolution dynamics of various regimes, which can give theoretic references for experimental laser design. However, traditional femtosecond mode-locked fiber laser modeling relies on the split-step Fourier method (SSFM) to iteratively resolve the nonlinear Schrödinger equation using a rather small step size,[5] which could cost plenty of time to simulate the laser output under different cavity settings. The low-efficiency SSFM method undoubtedly hinders the development of the inverse design and optimization of femtosecond lasers,[6][7] where numerous simulations are required.

In recent years, the combination of artificial intelligence (AI) and optics has drawn many concerns from the academic community.[8] AI techniques manifest extraordinary talents in equalization thereby improving the capacity of optics communication systems.[9][10] In laser optics, AI techniques substantially enhance the automatic optimization of mode-locked lasers[11]-[21] and the deep learning-based pulse characterization reveals the outstanding robustness against noise.[22][23] The modulation instabilities in optical fiber are analyzed by a fully-connected neural network[24] and precise phase retrieval in the interferometric fringe patterns is also achieved by AI models.[25],[26] On the other hand, with GPU-enabled compute unified device architecture (CUDA), the feed-forward computation of AI models can be substantially accelerated by parallel computing. Recently, the propagation inside optical fiber of high-order solitons and supercontinuum generation are simulated by AI techniques accurately, which is several orders of magnitude faster than the traditional SSFM method.[27] Then, a differential training approach is proposed to further reduce the computational complexity.[28]

However, it is a rather challenging task of modeling the femtosecond mode-locked fiber laser via AI techniques. First, compared to the soliton propagation and supercontinuum generation situations, the laser cavity is more complex due to the deeply-involved interplay among loss, gain, dispersion, nonlinearities, and other effects (e.g., saturate absorption). Second, unlike the previous research,[27],[28] the signal iteratively propagates inside the cavity for laser modeling. As a result, the impact of the aforementioned interplay on the ultimate laser output enhances as intra-cavity roundtrips increase, which further impedes the accurate femtosecond mode-locked fiber laser modeling using AI. Moreover, the imperative prior information of the laser settings (i.e., gain and cavity length) is crucial to ultimate laser output but is hidden from the AI models where the signal is usually taken as the only input.[27],[28] The prediction accuracy and the generalization ability of the AI model are bound to suffer a tremendous loss when missing the imperative prior information. The recent rise of physics-informed neural network (PINN) manifests itself as a method with superior performance in modeling nonlinear partial derivative equations[29],[30] and seems to be a good candidate for femtosecond mode-locked fiber laser modeling. Nevertheless, suffering from the strong bond between the PINN and the modeled physical equations, the generalization ability of the PINN is considerably confined.[30] Hence, the PINN is not suitable for femtosecond mode-locked fiber laser modeling where the generalization over various cavity settings is necessary. Until now, AI-based mode-locked fiber laser modeling still remains an open issue.

Here, a recurrent neural network (RNN) is used to realize fast and accurate femtosecond mode-locked fiber lasers modeling. In particular, to address the generalization problem of the AI model over various cavity settings, we propose a dimension-expansion-based prior information feeding method by which the prior cavity settings excluded by the signal can be appropriately fed to the AI model thereby assisting the model to infer. Our simulation results turn out that the proposed AI model with prior information feeding can complete the inference of 500 roundtrips less than 0.1 s on average with the acceleration of CUDA. Even on an identical CPU-based hardware platform, the proposed AI model is still 6 times faster than the traditional SSFM method. The generalization over the different cavity settings such as cavity length and gain is achieved by the RNN with prior information feeding. We hope our results could be a spark to inspire the extensive application of AI techniques in femtosecond laser modeling and other nonlinear optical systems.

## 2. Principles
### 2.1. Data generation via SSFM

The intra-cavity optical field evolution process is governed by the modified nonlinear Schrödinger equation as shown in Equation (1). In Equation (1), $T$ and $Z$ are respectively the time and distance variables, $A$ is the optical field envelope and $\omega_0$ is the central angular frequency. $\alpha$, $\beta_2$, $\beta_3$, $\gamma$ are the fiber loss, the second-order dispersion, the third-order-dispersion and nonlinear coefficient, respectively. $\Omega_g$ is the gain bandwidth. The gain process is described with the saturate gain $g = g_0/(1 + E_p/E_s)$, where $g_0$ is the small signal gain, $E_p$ is the pulse energy and $E_s$ is the saturate energy. The SSFM is a universal method to iteratively solve the nonlinear Schrödinger equation. The core idea of the SSFM is to split the fiber to propagate into numerous steps. For each step, only linear effects (i.e., dispersion, gain, and loss) or nonlinearities (i.e., self-phase modulation and self-steepening effect) are considered. The step size of the SSFM, the length of each step, is usually very small to ensure the modeling accuracy and the step size is 1 cm in our simulations. **Figure 1** shows the simulated femtosecond laser operating in the C band. The laser is pumped by a 0.3-m erbium-doped fiber (EDF), which is modeled by the saturate gain $g$. A piece of single mode fiber (SMF) is inserted to tune the cavity length. The saturate absorber (SA) acts as a mode locker with a modulation depth of 8% and saturable energy of 30 pJ. Subsequently, 10% of the energy is sent outside the cavity through a coupler. Note that the laser modeled by AI in this work operates under a traditional soliton regime with a SA as the mode locker. However, the proposed AI modeling method can also adapt to other types of mode-locked fiber lasers as long as the AI model is trained with the corresponding data.

$$\frac{\partial A}{\partial Z} + \frac{\alpha - g}{2} A + \frac{i\beta_2}{2} \frac{\partial^2 A}{\partial T^2} - \frac{\beta_3}{6} \frac{\partial^3 A}{\partial T^3} - \frac{g}{2\Omega_g^2} \frac{\partial^2 A}{\partial T^2} = i\gamma(|A|^2 A + \frac{i}{\omega_0} \frac{\partial(|A|^2 A)}{\partial T}) \quad (1)$$

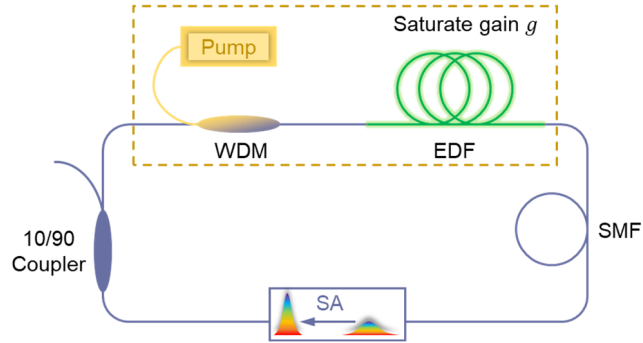

**Figure 1.** The simulated erbium-doped femtosecond laser. WDM, wavelength division multiplexer; EDF, erbium-doped fiber; SMF, single mode fiber; SA, saturate absorber

The simulation temporal window is 20 ps represented with 256 points. In the data generation via SSFM, the small signal gain $g_0$ is randomly selected in a range from 2 to 4.5, which covers the cases of failure to form a soliton, soliton formation, and soliton molecule formation. The length of the SMF inside the cavity is also randomly selected, therefore, the cavity length can range from 1.03 m to 2.05 m corresponding to a fundamental repetition rate ranging from 100 MHz to 200 MHz. To reduce the data amount, a fixed secant pulse with a duration of 5 ps is used as the initial state to accelerate the evolution process. The pulse seeding simulation method is common in the mode locked fiber laser modeling.[[31]-[33]] The max roundtrip number is fixed at 500 in simulations. Note that only one frame of data is recorded in each roundtrip manifesting a down-sampling operation in the frames. However, different from previous studies,[27][28] here the down-sampling rate is not fixed since the cavity length is changing thereby hampering the AI from learning the internal relation between frames. 1000 sets of data with random gain and fundamental repetition rates are generated via SSFM, where 960 sets are used for training, 20 sets are used for validation, and the rest 20 sets for the test.

## 2.2. The AI model and the error metric

**Figure 2a** shows the AI model structure for the femtosecond mode-locked fiber laser modeling. To retrieve the optical complex field with one neural network, the complex data of each roundtrip is unpacked before training thereby forming a 512-dimension vector, which is interleaved with real parts and imaginary parts. In previous studies where AI is used for high-order soliton propagation and supercontinuum generation,[27][28] the fiber channel is certain and all the information is contained in the frame data. Here, the first frame data is identical for different data sets. The evolution process is dominated by the small signal gain and cavity length, which are prior information excluded by the frame data. Therefore, a dense layer is used to complete the prior information feeding described in Equation (2), where $P$ is a 2-by-1 vector containing two prior parameters, $A$ is a 512-by-2 weight matrix and $b$ is the bias of the dense layer for prior information feeding. Then, a layer normalization operation ($LN$) is applied to the dimension-expansion result forming a 512-by-1 vector carrying prior information. Finally, the "prior vector" is broadcasted and added to the frame data $X$ with a dimension of 512-by-$W$ before the sequential layers, $W$ is the sequence length and the broadcasting add means each roundtrip in $X$ needs to add the "prior vector". As a result, the sequential data $X_P$ carrying both signal and prior information is obtained.

$$X_P = LN(A \cdot P + b) + X \tag{2}$$

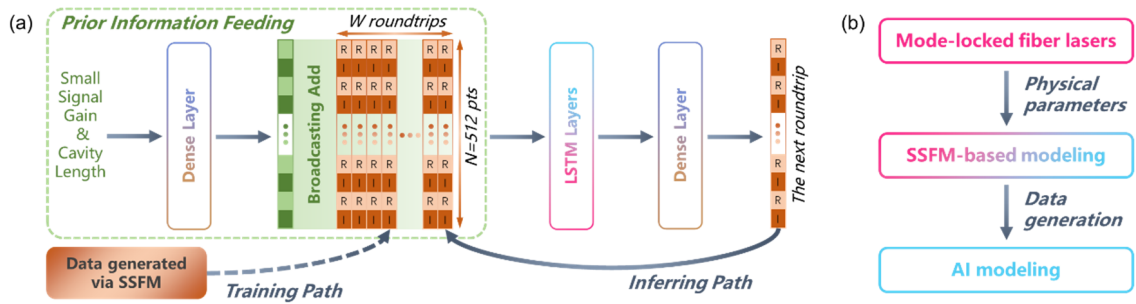

**Figure 2.** a) The AI model with prior information feeding for intra-cavity evolution process inference. SSFM, split-step Fourier method; LSTM, long-short term memory. b) The workflow for using the AI model.

Two unidirectional long-short term memory (LSTM) layers are used to capture the temporal relation along roundtrips and the sequence length is denoted as $W$ in Figure 2. Here, we choose a sequence length of 14. A dense layer summarizes the output of LSTM layers and predicts the next roundtrip. During training, the $W$ input roundtrips of LSTM layers stream from the generated via SSFM. However, when inferring, the last roundtrip in the $W$ input roundtrips is the previous prediction output by the model itself. The sequence of the input roundtrips functions like a sliding window. Note that we pad $W-1$ roundtrips, which are all identical to the first roundtrip, to form $W$ input roundtrips when predicting the second roundtrip. The workflow for using the AI model is shown in Figure 2b. The first step is to accurately model the real mode-locked fiber lasers using the theoretical SSFM-based modeling. Once the accurate SSFM-based modeling is realized, the modeling process can be substantially accelerated via training the AI model with numerous data generated by the SSFM model.

To quantitively evaluate the performance of the proposed AI model, a normalized root-mean-squared error (NRMSE) indicated in Equation (3) is used, where $D_{SSFM}$ is one real-and-imaginary-interleaved roundtrip generated by the SSFM and $D_{pred}$ is the corresponding AI prediction. Here, $m$ is the largest number of the real parts and imaginary parts among all the test datasets, $K$ denotes the number of test data sets ($K=20$ in this case), $R$ denotes the number of roundtrips in one data set ($R=500$ in this case), $N$ denotes the dimension of one real-and-imaginary-interleaved roundtrip ($N=512$ in this case).

$$\varepsilon_{NRMS} = \sqrt{\frac{\sum_{k,r,n}(D_{pred}-D_{SSFM})^2/m^2}{K \cdot R \cdot N}} \qquad (3)$$

## 3. Results
### 3.1. Soliton formation

Soliton is the product of the interplay between nonlinearities and anomalous dispersion. **Figure 3b** demonstrates the single soliton formation process in the spectral domain simulated by SSFM and the proposed AI model. The small signal gain $g_0$ is 2.85 and the cavity length reaches 1.4 m corresponding to a fundamental repetition rate of 146.7 MHz. It is obvious that the AI accurately predicts the soliton formation process including the spectral beating behavior before the spectrum settles down.[34][36] The full-field signals of the 100th roundtrip and the 500th roundtrip are respectively shown in Figure 3a and Figure 3c, and the ultimate pulse duration is 390 fs. It turns out that the full-field soliton formation process is well modeled by one AI model. Note that because the intensity on the sides is quite small, the phase on the sides is meaningless for both the temporal domain and the spectral domain. The soliton formation with different evolution dynamics and ultimate pulse duration is demonstrated in **Figure S1** in the supporting information.

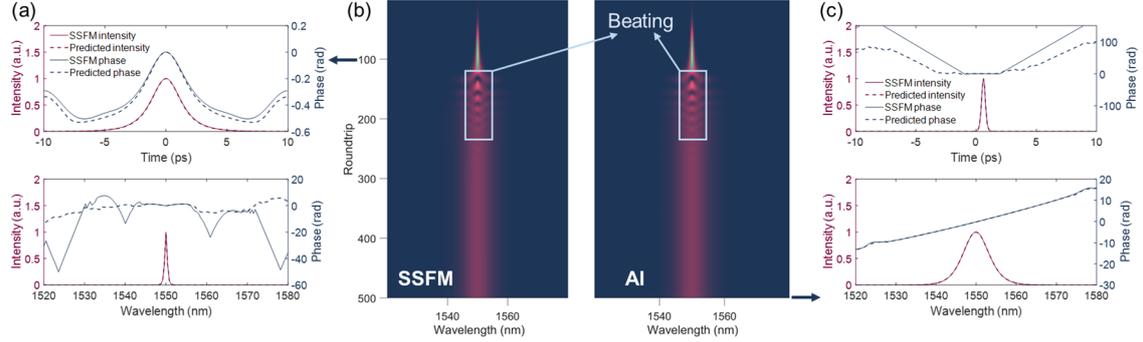

**Figure 3.** The soliton formation comparison between the SSFM and the AI model under a gain of 2.85 and a cavity length of 1.4 m. a) The full-field signal comparison at the 100th roundtrip. b) The spectral soliton formation process of the SSFM (left) and the AI model (right). c) The full-field signal comparison at the 500th roundtrip and the ultimate pulse duration is 390 fs. The SSFM-generated results are in solid lines while the AI predictions are in dashed lines in a) and c).

## 3.2. Soliton molecule formation

The laser generates a soliton molecule when further increasing the gain. The temporal soliton molecule formation processes under a gain of 3.96 simulated by SSFM and the proposed AI model are demonstrated in **Figure 4b**. The cavity length is 1.55 m and the corresponding fundamental repetition rate reaches 132.56 MHz. Because the evolution process is not converged at the 500th roundtrip yet (i.e., the inter-soliton separation $\tau$ and the inter-soliton relative phase $\phi$ is not settled at the 500th roundtrip as shown in Figure 4d), we further simulate 800 roundtrips and force the AI model to predict 800 roundtrips, which is beyond the training scope of 500 roundtrips. As shown in Figure 4b, the spectral evolution dynamics of soliton molecule formation is accurately retrieved by the AI model. Even in the roundtrips beyond the training scope (i.e., from the 500th roundtrip to the 800th roundtrip as highlighted by the dashed box in Figure 4b), the evolutions of spectral fringes are well predicted thereby manifesting the powerful generalization ability of the AI model over propagation distance. The maximum roundtrip number can be accurately predicted by the AI model varying from one cavity setup to another. Figure 4a and Figure 4c demonstrate the full-field signal comparisons at the 150th roundtrip and the 500th roundtrip, respectively. AI can both precisely emulate the symbolic fringe-pattern spectrum and the step-wise spectral phase of soliton molecule, proving the validity of full-field signal modeling via AI.

Moreover, the inter-soliton separation $\tau$ and the inter-soliton relative phase $\phi$ are extracted and their variations along roundtrips are shown in Figure 4d. The inter-soliton separation can be simply obtained by performing the Fourier transform on the spectrum.[35],[36] The inter-soliton relative phase ranging from 0 to $2\pi$ rad is extracted through a calculation involving inter-soliton separation, the central angular frequency, and the two wavelengths corresponding to the maximum two spectral peaks.[36] In Figure 4d, the top curves show the inter-soliton separation variation and the bottom curves show the inter-soliton relative phase variation. The entire evolution can be divided into three sections, which are the single-soliton section, the transient-soliton-molecule section, and the stable-soliton-molecule section.

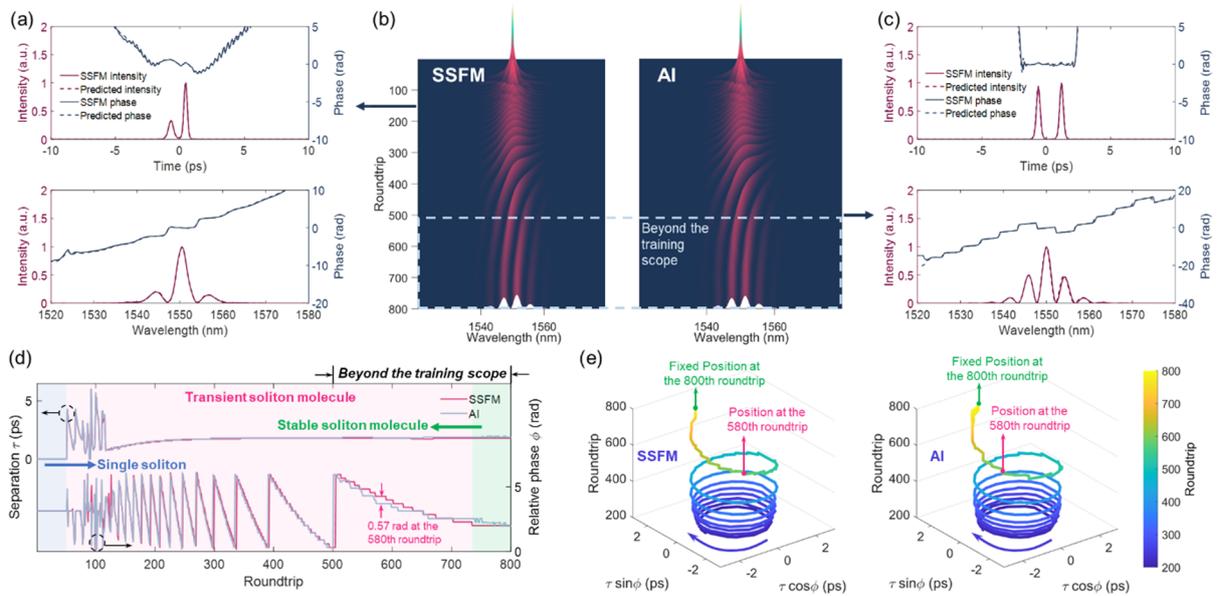

**Figure 4.** The soliton molecule formation comparison between the SSFM and the AI model under a gain of 3.96 and a cavity length of 1.55 m. a) The full-field signal comparison at the 150th roundtrip. b) The spectral soliton molecule formation process of the SSFM (left) and the AI model (right). c) The full-field signal comparison at the 500th roundtrip. d) variations of the inter-soliton separation $\tau$ (top curves) and the inter-soliton relative phase $\phi$ (bottom curves) along roundtrips and the ultimate inter-soliton separation is 1.8 ps. e) the clockwise evolutionary trajectories of the SSFM (left) and AI (right) in the interaction plane (the radius is the inter-soliton separation and the angle is the relative phase). The SSFM-generated results are in solid lines while the AI predictions are in dashed lines in a) and c). The SSFM-generated results are in rosy red while the AI predictions are in gray in d).

As shown in Figure 4d, at the beginning of the transient-soliton-molecule section, there is a period of chaotic evolution where the inter-soliton separation and relative phase are completely random. Then, an obvious transient soliton molecule is derived from the chaotic evolution (see Figure 4a) and the inter-soliton separation gradually rises till the stable soliton molecule is formed, whose inter-soliton separation and relative phase settle down to 1.8 ps and 2.01 rad, respectively. The predicted inter-soliton separation exhibits high consistency with the actual AI inter-soliton separation, even in the period of chaotic evolution. The same situation can also be found in the comparison between the actual and predicted inter-soliton relative phase. The regular step-wise shifts in the relative phase originate from the energy instabilities.[35] However, the mismatch between the actual and predicted inter-soliton relative phase magnifies as the roundtrip increases mainly due to the inherent error accumulation and propagation of the time series prediction. Concretely, the AI receives the correct data generated by the SSFM as the input in the training process. While during inferring, the previous prediction is feedback to the input for the next roundtrip inference. Thus, the error accumulates and propagates in the inferring process.

Figure 4e visually demonstrates the evolution process of the inter-soliton separation and the relative phase in the interaction plane, where the radius is the inter-soliton separation and the angle is the relative phase, from the 200th roundtrip to the 800th roundtrip, which is behind the period of chaotic evolution for better visual effects. Because both evolutionary trajectories rotate in a clockwise direction and settle down to two very close positions in the interaction plane as indicated by two green dots in Figure 4e. The evolutionary trajectories of the SSFM and AI manifest strong consistency and the residual inconsistency mainly streams from the relative phase mismatch in the zone where the roundtrip number is beyond the training scope. For instance, the predicting relative phase is 0.57 rad ahead of the actual relative phase at the 580th roundtrip as illustrated in Figure 4d. Since inter-soliton separations of the SSFM and the AI are very close when the roundtrip number is beyond the training scope, the phase overrun corresponds to a placement lead in the rotary evolutionary trajectory as illuminated in Figure 4e by comparing the positions of the two rosy red dots. As shown in Figure 4d, when the roundtrip number surpasses 500, the predicted relative phase observably overruns the actual relative phase for most roundtrips. As a result, the predicted evolutionary trajectory rotates faster averagely in the interaction plane. Nevertheless, the evolutionary trajectory of the SSFM first reaches the destination (i.e., the stable soliton molecule) since the relative phase of the SSFM comes from behind as shown in Figure 4d. Therefore, these pace differences result in slight differences between the two evolutionary trajectories. Overall, the proposed AI model successfully emulates the sophisticated evolution dynamics of the soliton molecule formation. **Figure S2** in the supporting information shows the soliton molecule formation with different evolution dynamics and the ultimate inter-soliton separation.

## 3.2. Failure of forming a soliton

When the gain is too small, the laser even fails to form a soliton. Given a gain of 2.07 and a cavity length of 1.97 m, the evolution dynamics simulated by SSFM and the proposed AI model are shown in **Figure 5b**. The AI is aware of the small gain via prior information feeding. Thus, it successfully predicts the final result of no-soliton formation. However, due to the small gain, the initial secant pulse becomes weaker along with the propagation and the intensity is comparable to noise. On the other hand, the aforementioned error accumulation and propagation problem of the time series prediction also contributes to prediction errors. Therefore, the AI cannot grasp the intensity and phase details as shown in Figure 5a and Figure 5c.

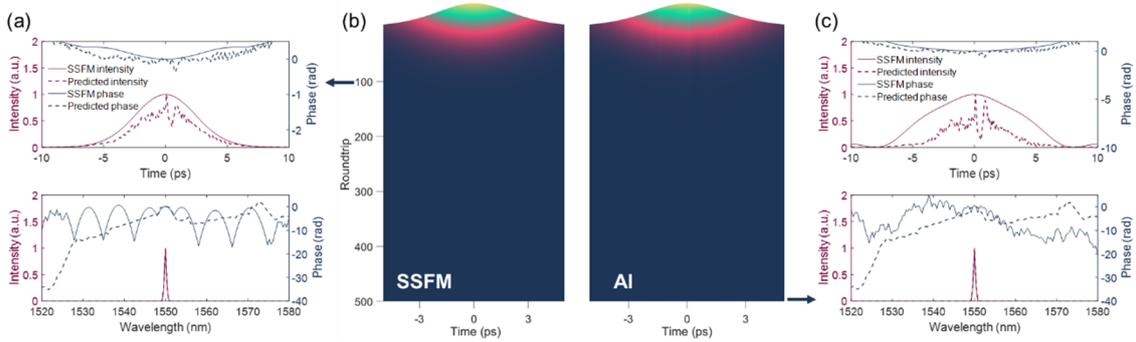

**Figure 5.** In the case of failing to form a soliton, the evolution dynamics comparison between the SSFM and the AI model under a small gain of 2.07 and a cavity length of 1.97 m. a) The full-field signal comparison at the 100th roundtrip. b) The temporal evolution dynamics of the SSFM (left) and the AI model (right). c) The full-field signal comparison at the 500th roundtrip. The SSFM-generated results are in solid lines while the AI predictions are in dashed lines in a) and c).

## 4. Discussions
### 4.1. Performance dependence on the number of LSTM layers

To investigate the performance dependence of the proposed AI model on the number of LSTM layers, 3 separate models with different numbers of LSTM layers are trained and tested under a fixed sequence length of $W = 10$. **Table 1** shows the comparison among these models in terms of the performance, computational complexity, model scale and simulation time. The model using only 1 LSTM layer performs worst with an NRMSE of 2.428 thereby failing to predict the evolution dynamics. After adding an LSTM layer, the NRMSE substantially reduces to 0.102 and the model can make an accurate prediction. However, further increasing the number of LSTM layers does not receive the performance improvement as expected. In virtue of the acceleration of CUDA, the 2-LSTM-layer model averagely consumes only 0.09 s to predict one data set with 500 roundtrips, which is ~146 times faster than the SSFM. Note that the SSFM calculation cannot be accelerated with CUDA. When testing on an identical AMD-CPU-based platform, the 2-LSTM-layer model is still ~6 times faster than the SSFM. On the other hand, less computational-intense convolutional layers are also applied to learning the temporal relations between sequential roundtrips but it turns out to be much worse than using LSTM layers. Accurate prediction with higher temporal resolution can be achieved by adding more hidden units in both the LSTM layers and the dense layer but it demands more computational resources for both the traditional SSFM method and the AI model.

**Table 1.** Comparison between AI models with different numbers of LSTM layers and the SSFM

|  | 1 LSTM layer | 2 LSTM layers | 3 LSTM layers | SSFM |
|---|---|---|---|---|
| NRMSE | 2.428 | 0.102 | 0.105 | N/A |
| FLOPs[a] | 6.360e7 | 1.477e8 | 2.317e8 | N/A |
| Number of parameters | 6.826e6 | 1.522e7 | 2.362e7 | N/A |
| Simulation time with CUDA[b] | 0.067 s | 0.090 s | 0.132 s | N/A |
| Simulation time[c] | 1.034 s | 2.106 s | 3.293 s | 13.145 s |

[a] Floating point operations; [b] The mean time of 200 simulations over an Nvidia RTX2080Ti GPU; [c] The mean time of 200 simulations over an AMD Ryzen 7 5800H CPU.

### 4.2. Performance dependence on the sequence length of LSTM layers

The sequence length is a critical hyper-parameter to sequence models. To evaluate the performance dependence of the proposed AI model on the sequence length, we train and test 4 separate models with different sequence lengths with 2 LSTM layers and the comparison results are shown in **Table 2**. All the AI models can accurately infer the evolution dynamics under various combinations of the gain and cavity length. As sequence length linearly increases, the computational complexity also linearly increases but the performance enhancement is trivial. The performance even drops when the sequence length increases from 14 to 18. In our simulations, the best sequence length is around 14.

**Table 2.** Comparison between AI models with different sequence lengths and the SSFM

|  | W=6 | W=10 | W=14 | W=18 | SSFM |
|---|---|---|---|---|---|
| NRMSE | 0.105 | 0.102 | 0.086 | 0.089 | N/A |
| FLOPs | 8.880e7 | 1.477e8 | 2.065e8 | 2.654e8 | N/A |
| Simulation time with CUDA[a] | 0.078 s | 0.090 s | 0.104 s | 0.134 s | N/A |
| Simulation time[b] | 1.366 s | 2.106 s | 2.899 s | 3.709 s | 13.145 s |

[a] The mean time of 200 simulations over an Nvidia RTX2080Ti GPU; [b] The mean time of 200 simulations over an AMD Ryzen 7 5800H CPU.

## 4.3. Current limitations

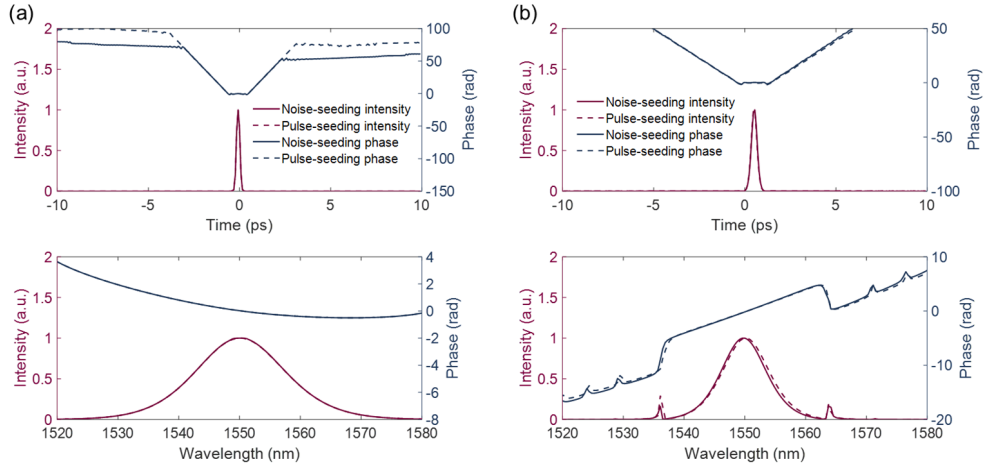

**Figure 6.** The stable lasing regimes comparison between the noise seeding method and the pulse seeding method. a) In a 200 MHz cavity. b) In a 50 MHz cavity. The noise seeding method in solid lines and the pulse seeding method in dashed lines.

Still, the proposed AI model remains two major limitations. First, the AI model cannot adapt to the dataset using the noise seeding method. However, given an identical cavity setting, both the noise seeding method and the pulse seeding method settle on nearly identical lasing regimes as shown in **Figure 6**, thereby manifesting the reasonability of using the pulse seeding method. On the other hand, it is common to use the pulse seeding method to accelerate the evolution process in the mode locked fiber laser.[[31]-[33]] Thus, stable lasing regimes can be achieved in fewer roundtrips thereby reducing the data amount.

Second, the AI model cannot adapt to longer cavities. The mode-locked fiber laser is a far more complex environment incorporating loss, gain, dispersion, nonlinearities, and other effects (e.g., saturate absorption). It is extremely challenging for AI to predict long-range light propagation in such a sophisticated environment. To accurately predict the 500-roundtrip evolution dynamics of a laser whose cavity length is 2 m corresponding to an FSR of ~100 MHz, the AI needs to accurately predict 1 km long-range light propagation, which is far longer than the previous studies.[27][28] One possible way to address the poor generalization over longer cavities is to increase the step size when using SSFM to produce data, by which the number of tiny steps separated by the step size of SSFM can be substantially reduced. However, in this way, the prediction precision improvement of the AI model comes at the cost of fineness reduction of the data generated by SSFM since the step size becomes larger. The two major limitations may be resolved by some advanced model, effective data extraction, and training methods, which we are going to study on in the near future.

## 5. Conclusion

To summarize, we demonstrate a completely data-driven method for modeling the femtosecond mode-locked fiber laser for the first time. Through the proposed prior information feeding, the AI model delivers robust generalization over the intra-cavity gain and cavity length. The AI model merely costs 0.09 s to predict one data set with 500 roundtrips with the acceleration of CUDA and the AI model is still 6 times faster than the SSFM even when testing on an identical CPU-based platform. We anticipate the superior running-time advantage of the AI model could substantially accelerate applications where numerous simulation data is required, for instance, the inverse design and optimization of femtosecond lasers.[6],[7] Further, we hope that the AI model could be adopted in other types of femtosecond lasers and becomes a general approach in the area of femtosecond laser modeling.

References

[1] J. Lee, Y. J. Kim, K. Lee, S. Lee, S. W. Kim, *Nat. Photonics* **2010**, *4*, 716.

[2] T. Udem, R. Holzwarth, T. W. Hänsch, *Nature* **2002**, *416*, 233.

[3] C. Xie, R. Meyer, L. Froehly, R. Giust, F. Courvoisier, *Light: Sci. & Appl.* **2021**, *10*, 126.

[4] M. Touil, S. Idlahcen, R. Becheker, D. Lebrun, C. Rozé, A. Hideur, T. Godin, *Light: Sci. & Appl.* **2022**, *11*, 66.

[5] X. Liu, D. Popa, N. Akhmediev, *Phy. Rev. Lett.* **2019**, *123*, 093901.

[6] A. Kokhanovskiy, E. Kuprikov, A. Bednyakova, I. Popkov, S. Smirnov, S. Turitsyn, *Sci. Rep.* **2021**, *11*, 13555.

[7] J. S. Feehan, S. R. Yoffe, E. Brunetti, M. Ryser, D. A. Jaroszynski, *Opt. Express* **2022**, *30*, 3455.

[8] G. Genty, L. Salmela, J. M. Dudley, D. Brunner, A. Kokhanovskiy, S. Kobtsev, S. K. Turitsyn, *Nat. Photonics* **2021**, *15*, 91.

[9] T. O'shea, J. Hoydis, *IEEE Transactions on Cognitive Communications and Networking* **2017**, *3*, 563.

[10] D. Wang, M. Zhang, *Frontiers in Communications and Networks* **2021**, *2*, 656786.

[11] U. Andral, R. Si Fodil, F. Amrani, F. Billard, E. Hertz, P. Grelu, *Optica* **2015**, *2*, 275.

[12] J. N. Kutz, S. L. Brunton, *Nanophotonics* **2015**, *4*, 459.

[13] R. I. Woodward, E. J. R. Kelleher, *Sci. Rep.* **2016**, *6*, 37616.

[14] D. G. Winters, M. S. Kirchner, S. J. Backus, H. C. Kapteyn, *Opt. Express* **2017**, *25*, 33216.

[15] R. I. Woodward, E. J. R. Kelleher, *Opt. Lett.* **2017**, *42*, 15.

[16] T. Baumeister, S. L. Brunton, J. N. Kutz, *JOSA B* **2018**, *35*, 617.

[17] G. Pu, L. Yi, L. Zhang, W. Hu, *Optica* **2019**, *6*, 362.

[18] A. Kokhanovskiy, A. Ivanenko, S. Kobtsev, S. Smirnov, S. Turitsyn, *Sci. Rep.* **2019**, *9*, 2916.

[19] X. Wei, J. C. Jing, Y. Shen, L. V. Wang, *Light: Sci. & Appl.* **2020**, *9*, 149.

[20] G. Pu, L. Yi, L. Zhang, C. Luo, Z. Li, W. Hu, *Light: Sci. & Appl.* **2020**, *9*, 13.

[21] Q. Yan, Q. Deng, J. Zhang, Y. Zhu, K. Yin, T. Li, D. Wu, T. Jiang, *Photonics Res.* **2021**, *9*, 1493.

[22] T. Zahavy, A. Dikopoltsev, D. Moss, G. I. Haham, O. Cohen, S. Mannor, M. Segev, *Optica* **2018**, *5*, 666.

[23] A. Kokhanovskiy, A. Bednyakova, E. Kuprikov, A. Ivanenko, M. Dyatlov, D. Lotkov, S. Kobtsev, S. Turitsyn, *Opt. Lett.* **2019**, *44*, 3410.

[24] M. Närhi, L. Salmela, J. Toivonen, C. Billet, J. M. Dudley, G. Genty, *Nat. Comm.* **2018**, *9*, 4923.

[25] S. Feng, Q. Chen, G. Gu, T. Tao, L. Zhang, Y. Hu, W. Yin, C. Zuo, *Advanced Photonics* **2019**, *1*, 025001.

[26] G. Pu, B. Jalali, *Opt. Express* **2021**, *29*, 20786.

[27] L. Salmela, N. Tsipinakis, A. Foi, C. Billet, J. M. Dudley, G. Genty, *Nat. Machine Intelligence* **2021**, *3*, 344.

[28] L. Salmela, M. Hary, M. Mabed, A. Foi, J. M. Dudley, G. Genty, *Opt. Lett.* **2022**, *47*, 802.

[29] M. Raissi, P. Perdikaris, G. E. Karniadakis, *Journal of Computational Physics* **2019**, *378*, 686.

[30] X. Jiang, D. Wang, Q. Fan, M. Zhang, C. Lu, A. P. T. Lau, *arXiv preprint* **2021**, *arXiv:2109.00526*.

[31] T. Schreiber, B. Ortaç, J. Limpert, A. Tünnermann, *Opt. Express* **2007**, 15, 8252.

[32] Z. Cheng, H. Li, P. Wang, *Opt. Express* **2015**, 23, 5972.

[33] H. Xu, X. Wan, Q. Ruan, R. Yang, T. Du, N. Chen, Z. Cai, Z. Luo, *IEEE Journal of Selected Topics in Quantum Electronics* **2017**, 24, 1100209.

[34] G. Herink, B. Jalali, C. Ropers, D. R. Solli, *Nat. Photonics* **2016**, *10*, 321.

[35] G. Herink, F. Kurtz, B. Jalali, D. R. Solli, and C. Ropers, *Science* **2017**, *356*, 6333.

[36] X. Liu, X. Yao, Y. Cui, *Phy. Rev. Lett.* **2018**, *121*, 023905.

# Supporting Information

**Fast predicting the complex nonlinear dynamics of mode-locked fiber laser by a recurrent neural network with prior information feeding**

*Guoqing Pu, Runmin Liu, Hang Yang, Yongxin Xu, Weisheng Hu, Minglie Hu, and Lilin Yi\**

G. Pu, H. Yang, Y. Xu, W. Hu, L. Yi
State Key Lab of Advanced Communication Systems and Networks, School of Electronic Information and Electrical Engineering, Shanghai Jiao Tong University, Shanghai, 200240, China
E-mail: lilinyi@sjtu.edu.cn

R. Liu, M. Hu
Ultrafast Laser Laboratory, Key Laboratory of Opto-electronic Information Science and Technology of Ministry of Education, College of Precision Instruments and Opto-electronics Engineering, Tianjin University, 300072 Tianjin, China

This document provides supporting information to "Fast predicting the complex nonlinear dynamics of mode-locked fiber laser by a recurrent neural network with prior information feeding", showing more results on the soliton formation and soliton molecule formation.

**1. Soliton formation with an ultimate pulse duration of 270 fs**

**Figure S1** shows temporal and spectral evolution dynamics of the soliton formation under a gain of 3.85 and a cavity length of 1.5 m. The gain of 3.85 is a critical gain between the single soliton formation and soliton molecules. Therefore, as shown in Figure S1a, there is a small pulse branching from the main soliton at about the 100th roundtrip. However, the gain is not large enough to support two stable solitons thereby the derived small pulse gradually fades away along with the propagation due to the fiber loss. Finally, only the main soliton is reserved and the ultimate pulse duration is 270 fs.

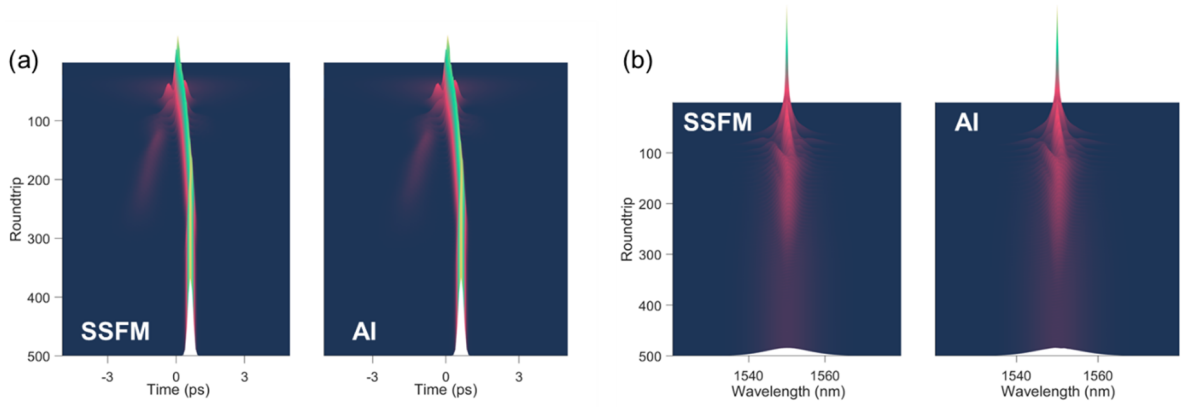

**Figure S1**. The soliton formation comparison between the SSFM and the AI model under a gain of 3.85 and a cavity length of 1.5 m. a) The temporal soliton formation process of the SSFM (left) and the AI model (right). b) The spectral soliton formation process of the SSFM (left) and the AI model (right).

**2. Soliton molecule formation with an ultimate inter-soliton separation of 1.09 ps**

Further increasing the gain, the laser will settle down to a state of soliton molecule. **Figure S2** shows temporal and spectral evolution dynamics of the soliton molecule formation under a gain of 4.46 and a cavity length of 1.06 m corresponding to a fundamental repetition rate reaching 194.14 MHz. As shown in Figure S2a, a soliton is firstly formed and then derives two solitons simultaneously. The inter-soliton separation varies as the roundtrip increases. The AI captures the general trend of the inter-soliton separation variation but the AI is unable to precisely track the inter-soliton separation variation. Anyway, as shown in Figure S2c, the AI accurately predicts the final roundtrip with an inter-soliton separation of 1.09 ps and the step-wise temporal phase is well retrieved. As shown in Figure S2b, the spectral beating along the roundtrips is observed, which is a necessary phase in the formation process of a soliton molecule.[1,2] Note that because the settled wavelength is not centered at 1550 nm, the temporal soliton position shifts along the roundtrips as shown in Figure S2a.

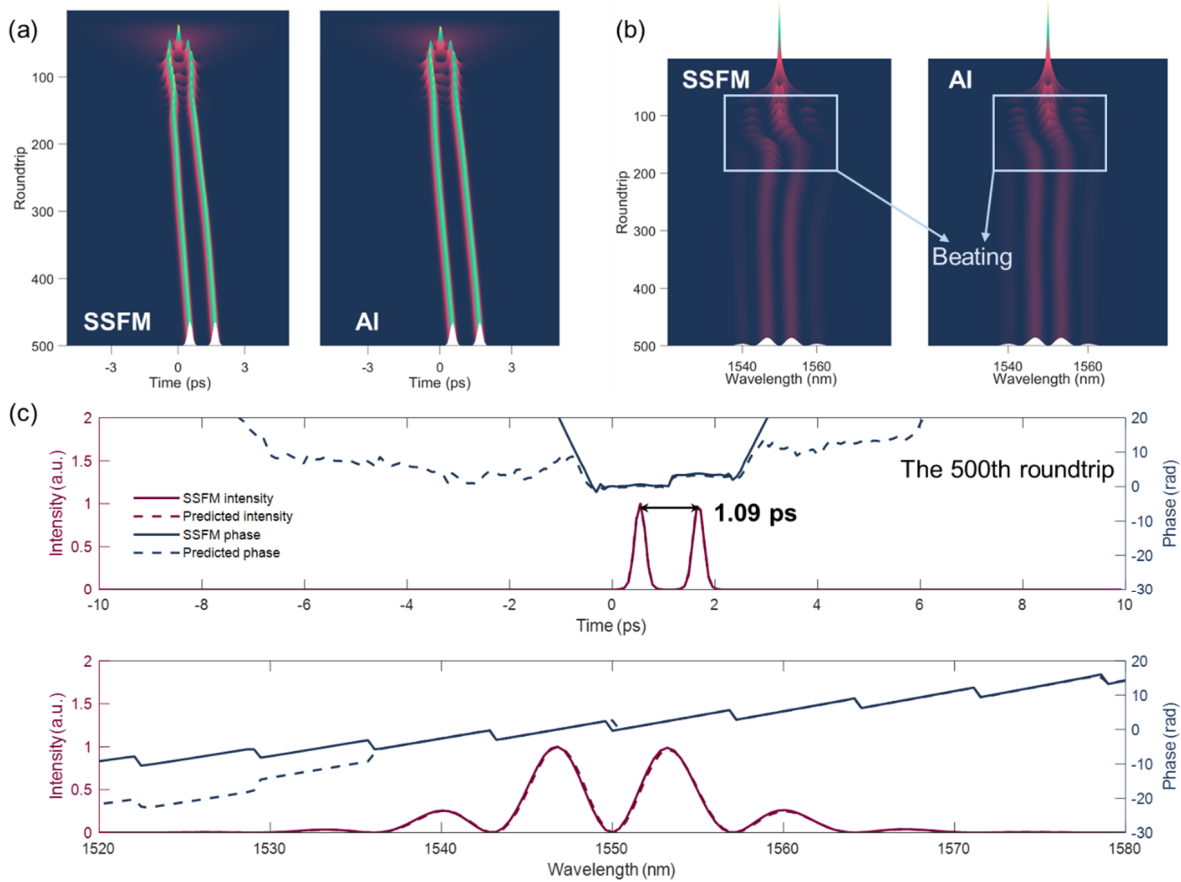

**Figure S2**. The soliton molecule formation comparison between the SSFM and the AI model under a gain of 4.46 and a cavity length of 1.06 m. a) The temporal soliton molecule formation process of the SSFM (left) and the AI model (right). b) The spectral soliton molecule formation process of the SSFM (left) and the AI model (right). c) The full-field signal comparison at the 500th roundtrip and the ultimate inter-soliton separation is 1.09 ps. The SSFM-generated results are in solid lines while the AI predictions are in dashed lines in c).

References
[1] G. Herink, F. Kurtz, B. Jalali, D. R. Solli, and C. Ropers, *Science* **2017**, *356*, 6333.
[2] X. Liu, X. Yao, Y. Cui, *Phy. Rev. Lett.* **2018**, *121*, 023905.